# Understanding the spatial burden of gender-based violence: Modelling patterns of violence in Nairobi, Kenya through geospatial information

Rina Friedberg[a], Clea Sarnquist[b], Gavin Nyairo[c], Mary Amuyunzu-Nyamongo[c], Michael Baiocchi[a,d]

*(a) Stanford Department of Statistics*
*(b) Stanford Department of Pediatrics*
*(c) African Institute for Health and Development*
*(d) Stanford Department of Epidemiology and Population Health*

**Abstract**

We present statistical techniques for analyzing global positioning system (GPS) data in order to understand, communicate about, and prevent patterns of violence. In this pilot study, participants in Nairobi, Kenya were asked to rate their safety at several locations, with the goal of predicting safety and learning important patterns. These approaches are meant to help articulate differences in experiences, fostering a discussion that will help communities identify issues and policymakers develop safer communities. A generalized linear mixed model incorporating spatial information taken from existing maps of Kibera showed significant predictors ($p < 0.05$) of perceived lack of safety included being alone and time of day; in debrief interviews, participants described feeling unsafe in spaces with hiding places, disease carrying animals, and dangerous individuals. This pilot study demonstrates promise for detecting spatial patterns of violence, which appear to be confirmed by actual rates of measured violence at schools. Several factors relevant to community building consistently predict perceived safety and emerge in participants' qualitative descriptions, telling a cohesive story about perceived safety and empowering communication to community stakeholders.

**Keywords**: spatial statistics; GIS; sexual assault; adolescents; violence prevention; sub-Saharan Africa

**Background**

*Introduction*

There is an ever-growing dataset describing the physical layout of our world, and with that comes new opportunities to model locations of violence in understudied communities. Among these communities are the informal settlements (colloquially, "slums") outside of Nairobi, Kenya, which suffer from high rates of violent crime (Ziraba et al., 2011) and gender-based violence (GBV) against children and adolescents (Baiocchi et al., 2019; Rosenman et al., 2019; Sarnquist et al., 2014; United Nations Children's Fund, 2012).

In a prior study (Rosenman et al., 2019), the authors modeled the self-reported rate of rape in various schools in settlements around Nairobi using univariate regressions on administrative features, such as student-to-teacher ratio, number of bathrooms, and building materials. With one exception, these features were not statistically significant at the 0.05 level, and none were strong predictors. The only statistically significant feature at the 0.05 level was the relative dropout rate, which captures the relative changes in gender proportion from class 5 to class 8, and even this was not a particularly strong predictor. Given the limited resources devoted to GBV prevention and the overall ineffectiveness of administrative features for predicted GBV rates, the authors cited the need for future research that can triage schools into groups with high and low risk of GBV in order to target prevention efforts in low-resource settings.

This current paper introduces a mixed-methods study design, which combines survey responses and open-ended interviews to describe safety in the informal settlements. This study design has several features to emphasize. First, as this is an exploratory, insight-generating study, the mixed-methods approach gives quantitative insights that pair with qualitative responses. Second, and related, we expect reduced demand effects with the quantitative data collection because the survey did not contain any questions about specific geographic features (for example: how do you feel around a bar?) rather, this study leverages maps and proximity to assess associations. In this design, the respondents can be potentially unaware or inarticulate of the correlates or causes of their concerns. Last, the link with maps enables researchers to construct any predictor available from

GPS data. Hence a fairly simple data collection procedure opens up doors to a rich variety of potential analyses.

The primary goal of this work is a proof of concept: to describe methods for safe but informative data collection and to develop principled statistical techniques for building GPS features and analyzing the data. More broadly, research like this can help give a voice to adolescents or children who may not normally not be included in conversation or have forums for communicating their needs. Following the social-ecological model of violence (Dahlberg & Krug, 2006), we look to leverage community-level data to generate insights that can be invisible on the individual-level of analysis.

The ultimate goal is to provide at-risk communities with tools to continue violence prevention work at the highest possible level: to understand the concerns of those in the community and then to build communities that are safer. We believe the use of maps and the connection to stories and drawings, backed by quantitative insights to analyze patterns of violence, will serve as a useful means for capturing and articulating needed changes in the community.

*Related Work*

Prior work has identified areas of San Diego county where reports of child maltreatment are concentrated (Barboza-Salerno, 2019), using geographically weighted regression models. Bayesian hierarchical methods were also used to identify spatiotemporal risk of child abuse in Los Angeles (Barboza-Salerno, 2019). Areas of high risk for domestic violence in João Pessoa, Brazil, were identified (Lucena et al., 2012) using the Getis-Ord index (Getis & Ord, 1992), which finds spatial hotspots for the outcome of interest. Hotspots can also be found with kernel density estimation, which uncovered locations of violent injuries in Campina Grande (Barbosa et al., 2019).

These can be effective for large datasets collected for non-GBV purposes, which are useful insights but are limited by the often-incidental variables collected about violent events. To our knowledge, there is very limited work demonstrating the usefulness of prospectively-collected, GPS-tagged survey data which can then be connected to existing spatial statistics to uncover locations related to violence. Moreover, a major contribution of this paper is linking geospatial data to specific features (such as the number of nearby bars). While only correlational analyses, these kinds of insights enable practitioners to interpret the results with a focus on neighborhood planning and policy implications, and creates enormous potential for leveraging existing geospatial data in global health applications.

**Methods**

*Procedure*

While this study was primarily a proof of concept, we oriented to three guiding questions: (i) Do feelings of safety vary by sex? (ii) How do feelings of safety change by day/night? (iii) What are the spatial predictors of variation in the feelings of safety? And do these spatial predictors change by sex and time-of-day? We were also interested in attempting to build prediction models that could help us suggest which schools are in locations that may feel more or less safe to these adolescents (i.e., using our model to predict out of sample locations) and to assess whether these predictions correlate with previously collected information on rates of sexual assault in these schools.

The measure of "safety" in this study is operationalized as self-reported feelings of safety. There are other candidate measures, such as crime statistics; these, however, may not give specific information about how adolescents feel. To accurately model perceived safety in different locations requires data about both safe and unsafe spaces. Logistical constraints in our setting prevent survey administrators from collecting data constantly throughout a neighborhood (e.g., mapping a continuous path through the community), and ethical constraints dictated that participants were not taken to any locations deemed to expose participants to risk. Moreover, to use a cell phone (or even a GPS tracker) throughout the course of a day would be risky to an adolescent in the informal settlements. Survey responses were therefore collected at seven pre-determined locations which satisfied these constraints but were anticipated to have diversity in responses to feelings of safety.

To minimize risk, the participants traveled in large groups with several escorts. While a given location may feel "unsafe" to a particular person when instructed to think about being in the space alone, the presence of additional travel companions reduces the risk profile. This aspect of the data collection procedure has some implications for the exact measurement obtained (e.g., likely to bias feelings of safety upward) but is well worth the benefit of mitigating risk.

Authors and collaborators at the African Institute of Health and Development (AIHD) selected seven locations in consultation with village elders, and recorded the locations' GPS coordinates before data collection. We chose some locations considered safe and some anticipated to feel less safe. So as not to expose the participants to increased risk locations, AIHD selected lower-risk locations near the sites where participants could go to identify the higher-risk location, but not be directly in that higher-risk location. Collaborators at AIHD recruited thirty-five adolescents, fifteen male and twenty female, to enroll in the study, and obtained written consent from parents and written assent from participants. This study was given ethical approval as a supplement to a clinical trial, registered at ClinicalTrials.gov #NCT02771132, by the KEMRI and Stanford Institutional Review Boards (IRBs).

*Data Collection*

On the morning of survey administration, study participants attended a briefing to explain the procedure and give them an opportunity to opt-out. Next, participants divided into groups separated by sex, to avoid social pressure on survey responses. AIHD collaborators took the participants to two or three of the selected locations at a time, before going to a convenient location where participants answered survey questions about the spaces they had just visited.

For each of the survey collection locations, participants answered a set of three questions, using a Likert scale from 1 (not at all) to 10 (very): (1) How safe do you feel at this location right now?; (2) How safe would you feel if you were alone?; (3) How safe would you feel at night? Survey administrators did not explain why any location was chosen; they did not preface by explaining whether or not it was expected to be safe, and did not point out any location features nearby. Correspondingly, *agreement* of quantitative and qualitative analyses provide certain kinds of insights to the researcher, and *disagreement* of these analyses provide different insights.

After data collection was complete at all seven locations, the study participants returned to where they received the training in the morning for a debrief session. Each participant chose one location they considered especially safe and one they considered unsafe, and described to survey administrators why they picked each spot. They also drew pictures of the locations, highlighting what made them feel how they described in each space.

*Feature Construction*

The GPS coordinates of each location provide a set of geographic features used to predict the participants' perceived safety using information from Map Kibera (mapkibera.org) and Open Street Map (OpenStreetMap contributors, 2015). These organizations offer public, crowd-sourced maps of neighborhoods like Kibera, which often have little to no map information from other resources ("Mapping Change: Community Information Empowerment in Kibera (Innovations Case Narrative: Map Kibera)," 2011). Many factors contribute to this, such as the frequent forced reorganization of buildings and communities inside the informal settlements ("Kenya slum demolished to make way for road," 2018). Individuals contributing to Map Kibera mark streets, rivers, churches, bars; but also street lights and water sources. We extracted this information using the R package Osmar (Eugster & Schlesinger, 2012).

We build concentric circles around each location, and counted the following within the circles. We considered circles of 50 meters (indicator for inclusion of a major river, i.e. Ngong or Motoine), 150 meters (street lights), and 400 meters (bars, churches, bus stops, and water sources). We selected these features in coordination with AIHD and using knowledge from prior experiences in these communities. Replication code, a tutorial, an example dataset, and a description for how we constructed features from Osmar are all available online at https://github.com/rinafriedberg/gbv-analytics/tree/master/Spatial_Statistics.

*Data Analysis*

Using both the map-derived features of the locations, along with participant-specific variables (sex), an indicator for circumstances being alone, and day or at night, we train a generalized linear mixed model (Nelder & Wedderburn, 1972) to predict perceived safety, using the R package lme4 (Bates et al., 2015).

The appropriate model here is an ANCOVA with repeated measures. As is standard for repeated measures, such as different survey responses from the same set of individuals, we include a random effect for each individual; correlation between responses from the same person is uninteresting for our study but still must be accounted for. We include fixed effects for the circumstances, gender, and map-derived features, which represent associations of interest where we expect significant correlation to be practically meaningful (Seltman, 2018).

We also include quantification of uncertainty: a bootstrap confidence interval, which requires careful construction that accounts for the quantities over which we want to measure error. The bootstrap standard errors build on the following set of assumptions. For every location, there is a true safety score (a grand mean), the quantity we want to estimate. If logistics and concern about risk exposure were not in play, we would randomly sample locations from the neighborhood, and randomly sample individuals from the target population, in order to construct a bootstrap dataset. This sampling frame allows for estimation of both the between-person variation but also for the between-location variation, which are important for better approximations of out-of-sample prediction error. A reasonable approximation, employed here, is to sample with replacement from both individuals and locations in our study sample. We repeat this 1000 times and train the model on these bootstrap datasets. We give 95% confidence intervals using quantiles from the corresponding results.

Last, we demonstrate the potential utility of this model by applying the model trained on survey results to predict safety of school locations in our sample. This is primarily an illustrative example, but serves to highlight the possible benefits of this method. We select 23 schools that are straightforward to find on Map Kibera and where we can confirm accurate GPS coordinates (i.e., a convenience sample). We then contrast patterns across predicted safety scores from the GPS model, with patterns from predicted rate of rape using school demographic features such as the number of participants, relative dropout rate, and the ratio of toilets and teachers per student. We then compare both models to the actual rates of rape collected during the randomized controlled trial for these 23 schools. We compare boxplots of these results on a partition of the dataset into five groups based on observed rate of rape in Figure 2.

**Results**

*Descriptive Statistics*

Figure 1 displays boxplots of the survey responses for perceived safety at the time of data collection, and at night. Matching intuition, the safety scores are lower at night for every location, even for relatively safe locations such as six and seven. Female participants reported slightly higher overall safety scores at locations three and five, and at locations six and seven all participants reported high perceived safety.

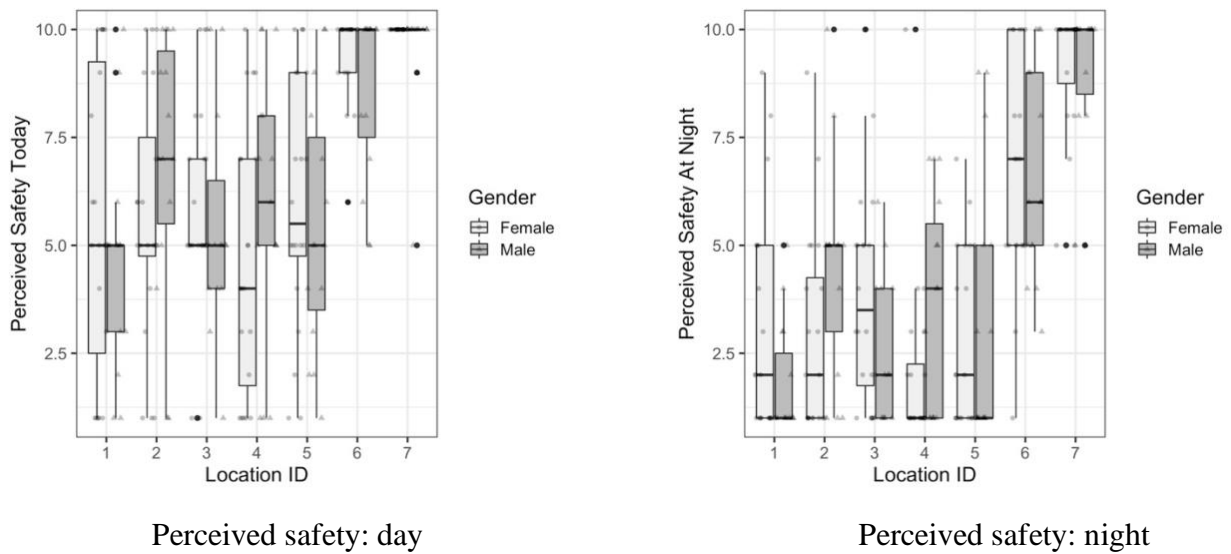

Perceived safety: day    Perceived safety: night

Figure 1. Boxplots displaying self-reported safety at each of the seven locations, stratified by gender. The left panel shows perceived safety at the time of study collection, and the right panel perceived safety at night.

*Quantitative modeling results*

Table 1 displays coefficients and bootstrap uncertainty measures from LMER, from the model predicting safety score from GPS features and circumstances (alone or at night). Factors that are associated with significantly lower safety reports are the circumstances of being alone (-2.29; 95% CI (-2.94, -1.49)) and being at night (-2.12; 95% CI (-2.94, -1.48)). The other coefficients have higher bootstrap standard errors. Among those, features that raised reported safety scores were the number of nearby bars and nearby water sources; features that lowered safety scores were the number of nearby lights, bus stops, and religious buildings, and a nearby river. Among some coefficients, there is relatively inconclusive data; for example, the number of nearby bars was between 3-5 for every location except one, which had 11 nearby bars.

Replications studies – with more locations and study participants included – would show whether these features are truly insignificant, or if this is due to limited statistical power in a small sample size. Table 2 (in the appendix) shows a similar model, but including interactions of all features with sex. Those interactions would help answer the question of if the factors which are predictive of feelings of safety vary by sex. We found no statistically significant associations – which is unsurprising given this study was underpowered for estimating these interactions.

| Feature | Coefficient | 95% Bootstrap confidence Interval | Bootstrap standard error |
|---|---|---|---|
| Male | 0.08 | (-0.64, 1.34) | 0.38 |
| Alone | -2.29 | (-2.94, -1.49) | 0.38 |
| Night | -2.12 | (-2.94, -1.48) | 0.50 |
| Number of lights within 150 meters | -5.63 | (-13.19, 6.92) | 4.33 |
| Number of bars within 400 meters | 8.81 | (-8.61, 25.04) | 7.39 |
| Number of bus stops within 400 meters | -15.15 | (-21.08, 27.39) | 9.67 |
| Number of water sources within 400 meters | 1.21 | (-0.62, 1.38) | 0.51 |
| Number of religious buildings within 400 meters | -1.99 | (-2.74, 1.34) | 0.97 |
| Less than 50 meters from Ngong or Motoine rivers | -19.27 | (-37.87, 61.44) | 18.83 |

Table 1. Coefficients and bootstrap confidence intervals for predicting safety scores.

*Predicting School Safety*

The panels of Figure 2 show two predictive models of safety scores for 23 schools. The x-axis of the boxplots are divided into five groups, the same x-axis is used in both panels. The x-axis divides the schools into equal quintiles in terms of their actual reported rate of rape. In the left panel is the spatial model, using the model fit on the participants and then predicting using geographic features collected for the 23 schools (i.e., these are "out-of-sample" predictions). In the right panel, we show predicted rate of rape using the observed rates of rape and the school demographic features (this is the model fit in (Rosenman et al., 2019)); these y-axis values are not comparable.

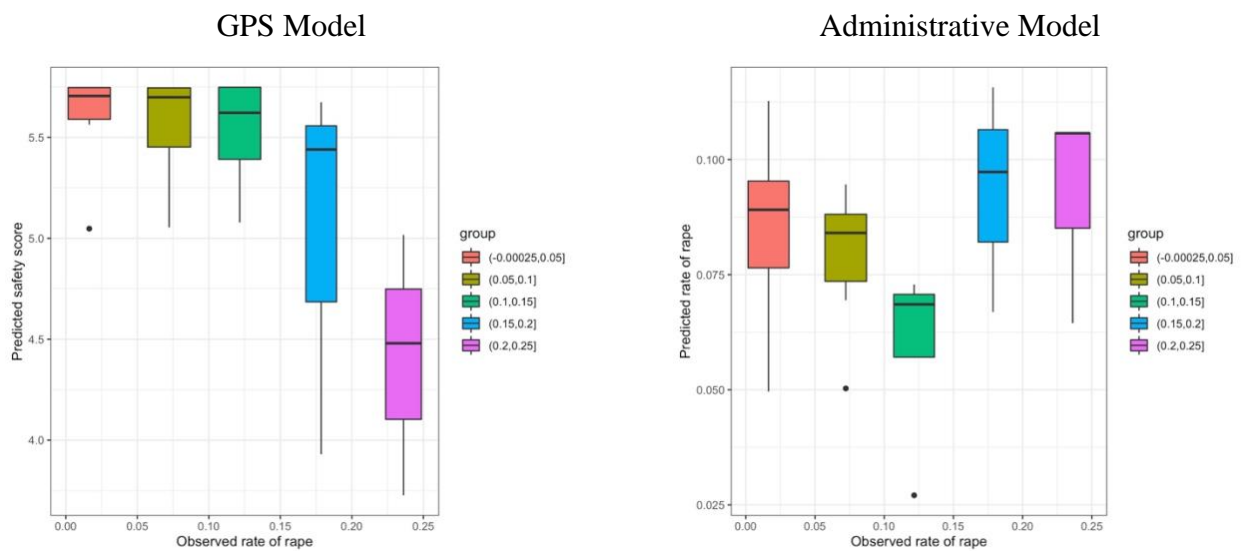

Figure 2. Boxplots of (left) predicted safety score and (right) predicted rate of rape on the y-axis, by the observed rate of rape on the x-axis, for 23 schools.

We observe that using GPS features, the safety scores have a steadily downward trend, monotonic in the observed rate of rape. On the other hand, there is a more U-shaped pattern using the school demographic features to predict reported rate of rape, indicating that on this subset of schools, the school features are not linearly correlated with the observed rate of rape.

*Qualitative analysis and synthesis with quantitative*

Participants reported that locations less than 50 meters from major rivers were less safe than other locations (coefficient from spatial model: -19.27). Several candidate reasons emerge from the interviews. Many participants mention that locations with hiding spots, especially trees and ditches that are near major rivers, are very unsafe, because "some robbers, kidnappers, or even murderers may hide in the tree waiting for one to pas [sic] so that they kidnap you. For girls, they may rape you or even kill you." Rivers can also pose physical safety concerns; one participant mentioned a broken bridge, which "is very unsafe because even a child can loose [sic] his steps and fall inside the river, the child might loose his life."

Another trend throughout the quantitative and qualitative responses is the impact of adults in the area. Thirty out of 35, or 86%, of children chose location seven, which village elders marked as safe due to the high concentration of chiefs and police in the area, as the most safe location. Participants explained that this location is very safe "because there are watchmen all over and there are some police walking," and because "there are special people who do guidance and counseling and there are no crimes or civil war taking place." On the other hand, another participant chose the least safe location because "some people are drug dealers," "people are fighting," and "grown up men seducing little or young girls, they are impregnating them."

**Discussion**

We analyzed participants' perceived safety at seven pre-determined locations throughout Kibera. We trained a repeated measures regression model with random effects for the individual participants and fixed effects for being alone, being at night, number of very nearby (within 150 meters) lights, number of nearby (within 400 meters) bars, bus stops, water sources, and religious buildings, and whether the location is less than 50 meters from the Ngong or Motoine rivers. Significant features at the p=0.05 level were the circumstances of being alone or at night, both of which lowered perceived safety.

Among features not statistically significant, the number of lights, bus stops, and religious buildings, and close distance to the rivers had negative coefficients (which represent lowered safety scores). From prior discussions with participants, we know transportation hubs have been discussed as being unsafe so this hints that our study was not perfectly aligned with intuitive understandings of safety. The direction of causality is obviously not established by type of observational study: for example, we should consider whether lights and churches are added to the least safe locations in order to mitigate existing safety issues, and therefore whether high presence of street lights and churches do not cause lower safety responses, but indicate that adults in power also expected those locations to be unsafe. Again, this is a pilot study so issues of power should constrain interpretations.

The number of nearby bars and water sources raised safety scores. The result about the number of bars was surprising; we would have anticipated bars to predict feelings of unsafety, due to the increased presence of drunk adults nearby. A possible explanation is that in a dataset with only 7 locations, outliers can completely dictate the results of the model coefficient. In a future study, we would intentionally choose locations with more variation in the number of bars (perhaps assigning two locations to each: 0, 3-5, and >10 bars), in order to determine whether this finding is spurious or meaningful.

Several parallels emerged between the quantitative results and participant interviews, despite not asking the participants about the locations' proximities to bars, churches, etc. Proximity information was implicit in each location, and the GPS analysis brought it to light. This is a feature of this study design we find particularly of interest: that measurements implicit (pulled from maps) can be put in relation to the open-responses from the respondents. It is also important to note that the quantitative questions are non-directive – that is, we ask "how safe do you feel?" with no reference to specifics of the location – and then the variation in that response is attributed to implicitly collected covariates from online maps. Possibly due to the small sample size, the coefficients discussed here are not statistically significant; that said, we can still discuss their directions and possible connections to participant responses. Future research with a greater sample size would be valuable in validating or calling into question this exploratory discussion.

Perhaps most suggestive of this approach's utility: On a subset of 23 schools, the GPS features detected patterns in the reported rate of rape that school administrative/demographics alone could not successfully model. This use of the spatial model is especially exploratory; one might expect that the schools easy to find and confirm on online map resources are qualitatively different from the full set of schools. We do verify that the average rate of rape is not significantly different among this subset of schools; the average reported rate of rape across these schools was 8.5% (95% bootstrap confidence interval on 1000 replications (5.8%, 11.6%)), up from the overall average of 7.2%, but this average is contained in the confidence interval. Despite the likelihood of other differences between these schools, we can still observe the potential of GPS features to improve our ability to triage which schools are particularly unsafe for participants, to understand specifically why, and to inform policies and prevention strategies aiming to improve safety. It might be surprising to see that a predictive model using school-specific administrative measures of a school were less predictive of the rate of rape within the school than a predictive model using just spatial variables. While this study certainly does not prove that spatial models are better at this prediction task - for example there may be other administrative records that would have been informative - it merits further exploration.

We can frame these developments in terms of the social-ecological model (Dahlberg & Krug, 2006), which describes patterns of violence in increasing terms of the individual, interpersonal relationships, community, and society. We leverage information about the community in which these participants live, in order to learn about where violence happens and provide tools for building safer spaces for participants in the future.

**Limitations**

This study is a proof of concept, and in order to verify results and gain statistical power, should be repeated with more study participants and locations. More specific hypotheses, particularly about spatial features (e.g., density of bars) would be useful for designing the study by selecting locations which have more variation in the covariates of most interest. It would also be worthwhile to repeat the study in different informal settlements around Nairobi and in other regions, in order to collect data more representative of the whole population.

Moreover, safety concerns necessitate a somewhat limited data collection process. We cannot expose participants to actual risk. Careful thought about how to ensure safety (e.g., using more resources to secure the participants and thus allow access to areas) or to get participants in the right "headspace" to elicit responses about their perception of a location (e.g., detailed videos). It is possible that these concerns about safety cannot be completely circumvented, but it should be acknowledged when considering the results.

It is also important to recognize that maps (particularly in a highly dynamic communities like Kibera) may be out of date by the time we collect our survey data. That is, there are certain kinds of measurement error introduced by using existing maps.

The community-specific insights generated by this study, and the potential to use stories and the drawings from the adolescents to communicate concerns, is exciting. In particular, we imagine the integrating these insights onto a map that all community members would have access to in order to understand perceived risk. But this also introduces serious concerns about nefarious-actors. It is conceivable that the specificity of some of the insights generated by this kind of study design would lead some actors gaining information that could be used to harm or exploit members of the community. Before more work can be done in producing highly-integrated outputs from this kind of study design, we need to work out a means for communicating actionable insights while obscuring any "vulnerabilities" that might be localized to a particular location or subgroup in the community.

**Conclusions**

In this paper, we created a small set of features to demonstrate potential. This study provides insights into how existing, publicly-available spatial data, in this case from online maps, can be used to identify and target policy and prevention in a low-resource setting. A fairly small set of easy-to-access covariates provided information about safety and danger in these communities that was very similar to that from an intensive data-collection procedure with human subjects. Furthermore, these covariates predicted risk of sexual assault in some schools in these communities, which could facilitate targeting of limited resources to areas of likely high need.

With a simple survey and GPS coordinates, we could create as many covariates as are interesting or useful, and then use variable selection to trim down the model. This allows for an enormous variety of models and of potential insights, and is an opportunity for creative research throughout violence prevention work to quantitatively test insights about space, violence, and underserved communities. Through this project and the doors it opens, we can provide a voice to individuals who do not always have an opportunity to advocate for their own safety, and generate myriad insights about how to better integrate their voices in program and policy decisions.

**Acknowledgments:** This work was funded by the Marjorie Lozoff Prize from the Stanford Clayman Institute for Gender Studies, awarded to Rina Friedberg, with supplemental funds from the George Rosenkranz Prize, awarded to Michael Baiocchi. The authors are grateful for helpful discussions with Judy Engleman and Barbara Brody, and would like to thank the Stanford Geo-Spatial Center and Stace D. Maples for incredibly helpful guidance and resources.

**References**

Baiocchi, M., Friedberg, R., Rosenman, E., Amuyunzu-Nyamongo, M., Oguda, G., Otieno, D., & Sarnquist, C. (2019). Prevalence and risk factors for sexual assault among class 6 female students in unplanned settlements of Nairobi, Kenya: Baseline analysis from the IMPower & Sources of Strength cluster randomized controlled trial. *PloS One*, *14*(6), e0213359–e0213359. PubMed. https://doi.org/10.1371/journal.pone.0213359

Barbosa, K. G. N., Walker, B. B., Schuurman, N., Cavalcanti, S. D. L. B., Ferreira E Ferreira, E., & Ferreira, R. C. (2019). Epidemiological and spatial characteristics of interpersonal physical violence in a Brazilian city: A comparative study of violent injury hotspots in familial versus non-familial settings, 2012-2014. *PloS One*, *14*(1), e0208304–e0208304. PubMed. https://doi.org/10.1371/journal.pone.0208304

Barboza-Salerno, G. (2019). Examining Spatial Regimes of Child Maltreatment Allegations in a Social Vulnerability Framework. *Child Maltreatment*, *25*, 107755951985034. https://doi.org/10.1177/1077559519850340

Bates, D., Mächler, M., Bolker, B., & Walker, S. (2015). Fitting Linear Mixed-Effects Models Using lme4. *Journal of Statistical Software; Vol 1, Issue 1 (2015)*. https://doi.org/10.18637/jss.v067.i01

Dahlberg, L. L., & Krug, E. G. (2006). Violence a global public health problem. *Ciência & Saúde Coletiva*, *11*, 277–292.

Eugster, M. J. A., & Schlesinger, T. (2012). *osmar: OpenStreetMap and R*. R Journal. Accepted for publication on 2012-08-14. http://osmar.r-forge.r-project.org/RJpreprint.pdf.

Getis, A., & Ord, J. K. (1992). The Analysis of Spatial Association by Use of Distance Statistics. *Geographical Analysis*, *24*(3), 189–206. https://doi.org/10.1111/j.1538-4632.1992.tb00261.x

Kenya slum demolished to make way for road. (2018, July 23). *BBC News*. https://www.bbc.com/news/world-africa-44931808

Lucena, K. D. T. de, Silva, A. T. M. C. da, Moraes, R. M. de, Silva, C. C. da, & Bezerra, I. M. P. (2012). Análise espacial da violência doméstica contra a mulher entre os anos de 2002 e 2005 em João Pessoa, Paraíba, Brasil. *Cadernos de Saúde Pública*, *28*, 1111–1121.

Mapping Change: Community Information Empowerment in Kibera (Innovations Case Narrative: Map Kibera). (2011). *Innovations: Technology, Governance, Globalization*, *6*, 69–94. https://doi.org/10.1162/INOV_a_00059

Nelder, J. A., & Wedderburn, R. W. M. (1972). Generalized Linear Models. *Journal of the Royal Statistical Society. Series A (General)*, *135*(3), 370–384. JSTOR. https://doi.org/10.2307/2344614

OpenStreetMap contributors. (2015). *Planet dump [Data file from 12/30/2019]. Retrieved from https://planet.openstreetmap.org*.

Rosenman, E., Sarnquist, C., Friedberg, R., Amuyunzu-Nyamongo, M., Oguda, G., Otieno, D., & Baiocchi, M. (2019). Empirical Insights for Improving Sexual Assault Prevention:

Early Evidence from a Cluster-Randomized Trial of IMPower and Sources of Strength. *Violence Against Women*. https://doi.org/10.1177/1077801219886380

Sarnquist, C., Omondi, B., Sinclair, J., Gitau, C., Paiva, L., Mulinge, M., Cornfield, D. N., & Maldonado, Y. (2014). Rape Prevention Through Empowerment of Adolescent Girls. *Pediatrics*, *133*(5), e1226. https://doi.org/10.1542/peds.2013-3414

Seltman, H. J. (2018). *Experimental Design and Analysis*.

United Nations Children's Fund. (2012). *Violence against Children in Kenya: Findings from a 2010 National Survey. Summary Report on the Prevalence of Sexual, Physical and Emotional Violence, Context of Sexual Violence, and Health and Behavioral Consequences of Violence Experienced in Childhood Nairobi, Kenya: United Nations Children's Fund, Prevention DoV*.

Ziraba, A. K., Kyobutungi, C., & Zulu, E. M. (2011). Fatal injuries in the slums of Nairobi and their risk factors: Results from a matched case-control study. *Journal of Urban Health : Bulletin of the New York Academy of Medicine*, *88 Suppl 2*(Suppl 2), S256–S265. PubMed. https://doi.org/10.1007/s11524-011-9580-7

**Appendix**

| Feature | Coefficient | 95% Bootstrap confidence Interval | Bootstrap standard error |
|---|---|---|---|
| Male | 4.47 | (-0.70, 1.24) | 0.48 |
| Alone | -2.35 | (-3.01, -1.48) | 0.38 |
| Night | -2.21 | (-2.98, -1.48) | 0.50 |
| Number of lights within 150 meters | -5.34 | (-5.52, 14.43) | 7.57 |
| Number of bars within 400 meters | 8.46 | (-6.00, 8.83) | 3.88 |
| Number of bus stops within 400 meters | -13.80 | (-22.64, 24.77) | 16.51 |
| Number of water sources within 400 meters | 1.21 | (-0.75, 0.93) | 1.03 |
| Number of religious buildings within 400 meters | -1.91 | (-1.05, 1.41) | 1.67 |
| Less than 50 meters from Ngong or Motoine rivers | -18.92 | (-20.36, 12.26) | 8.97 |
| Male x Alone | 0.15 | (-1.99, 1.29) | 0.81 |
| Male x Night | 0.22 | (-2.52, 1.23) | 0.95 |
| Male x Number of lights within 150 meters | -0.69 | (-0.66, 2.63) | 0.85 |
| Male x Number of bars within 400 meters | 0.82 | (-1.79, 4.58) | 1.54 |
| Male x Number of bus stops within 400 meters | -3.13 | (-17.67, 3.21) | 5.23 |
| Male x Number of water sources within 400 meters | -0.19 | (-0.31, 0.15) | 0.10 |
| Male x Number of religious buildings within 400 meters | -0.82 | (-0.19, 2.85) | 0.90 |
| Male x Less than 50 meters from Ngong or Motoine rivers | 4.47 | (-8.40, 3.05) | 2.82 |

Table 2. Coefficients and bootstrap confidence intervals for predicting safety scores, with interactions between covariates and sex included.